\documentclass[usenatbib]{mn2e}
\usepackage{times}
\usepackage{epsfig}
\newcommand{\be}{\begin{equation}}
\newcommand{\ee}{\end{equation}}
\newcommand{\etal}{et al.\ }
\newcommand{\chandra}{{\it Chandra} }
\newcommand{\xmm}{{\it XMM-Newton} }
\newcommand{\rosat}{{\it ROSAT} }

\title[Scaling relations in fossil groups]
{Scaling relations in fossil galaxy groups}
\author[Khosroshahi \etal]{
Habib G. Khosroshahi\thanks{E-mail:
habib@star.sr.bham.ac.uk},
 Trevor J. Ponman \& Laurence R. Jones \\
School of Physics and Astronomy, The University of Birmingham,
Birmingham B15 2TT, UK}

\begin{document}

\date{Accepted, Received}

\pagerange{\pageref{firstpage}--\pageref{lastpage}} \pubyear{2006}

\maketitle

\label{firstpage}

\begin{abstract} 
Using \chandra X-ray observations and optical imaging and spectroscopy
of a flux-limited sample of 5 fossil groups, supplemented by 
additional systems from the literature, we provide the first detailed
study of the scaling properties of fossils compared to normal groups
and clusters. Fossil groups are dominated by a single giant
elliptical galaxy at the centre of an extended bright X-ray halo. In
general, all the fossils we study show regular and symmetric X-ray emission,
indicating an absence of recent major group mergers. We study the
scaling relations involving total gravitational mass, X-ray
temperature, X-ray luminosity, group velocity dispersion and the
optical luminosity of the fossil groups.

We confirm that, for a given optical luminosity of the group, fossils
are more X-ray luminous than non-fossil groups. Fossils, however, 
fall comfortably on the conventional $L_X-T_X$ relation
of galaxy groups and clusters, suggesting that their X-ray luminosity and 
their gas temperature are both boosted, arguably, as a result
of their early formation. This is supported by other scaling relations
including the $L_X-\sigma$ and
$T_X-\sigma$ relations in which fossils show higher X-ray luminosity and 
temperature for a given group velocity dispersion. We find that 
mass concentration in fossils is higher than in non-fossil groups and 
clusters. In addition, the $M_X-T_X$ relation suggests that fossils are 
hotter, for a given total gravitational mass, both consistent with 
an early formation epoch for fossils. We show that the mass-to-light ratio 
in fossils is rather high but not exceptional, compared to galaxy groups 
and clusters. The entropy of the gas in low mass fossils appears to be 
systematically lower than that in normal groups, which may explain why 
the properties of fossils are more consistent with an extension of 
cluster properties. We discuss possible reasons for this difference 
in fossil properties and conclude that the cuspy potential raises the 
luminosity and temperature of the IGM in fossils. However, this works
in conjunction with lower gas entropy, which may arise from 
less effective preheating of the gas.

\end{abstract}


\section{Introduction}

Galaxy groups are key systems in advancing our
understanding of structure formation and evolution. They contain
the majority of galaxies in the universe, and are precursors to the most
massive structures, i.e. clusters, giving them cosmological importance.
However, they show departures from the scaling relations obeyed by
galaxy clusters indicating that they are not simply scaled-down versions
of clusters.

Theoretical or computational models (e.g. \citet{nfw95})
based on simple gravitational collapse and shock
heating would lead to self-similar structure of the inter-galactic medium 
(IGM) in clusters and groups. This in turn implies scaling relations
between the global properties of clusters:
$L_X\propto T_X^2$, $L_X\propto \sigma^4$ and $M\propto T_X^{3/2}$,
where $L_X$ is the total X-ray luminosity, $T_X$ the gas
temperature, $\sigma$ the velocity dispersion of cluster galaxies
and $M$ is the total gravitational mass of the cluster.
While some studies find that the 
observed properties of the most massive clusters are close to the above
relations \citep{allen98,xuejinwu01}, the self-similar model clearly breaks 
down in smaller systems,
with observations indicating lower than expected luminosities for a
given temperature or velocity dispersion. \citet{wjf97}
found $L_X\propto T^3$ for clusters observed with the {\it Einstein} 
X-ray Observatory. For galaxy groups, \citet{mz98} found $L_X
\propto T^3$, consistent with clusters, while \citet{helsdon00}
and \citet{xue00} found much steeper relations, $L_X\propto T^5$.

Galaxy groups are rapidly evolving and diverse systems, and many are not
virialised (eg. \citet{jesper06}). Thus studying a sample of 
well-characterised galaxy groups, in terms of their stellar properties and
IGM, might help us to understand some of the observed diversity 
in group properties.

\citet{jones03} studied a flux-limited sample of old 
galaxy groups known as ``fossil groups'' and 
found significant differences in their intergalactic hot gas properties 
in comparison to other galaxy groups. By estimating the space density
of fossil groups, they showed that fossils are as numerous as galaxy 
clusters. Fossil groups are particularly
important systems to study, in the context of scaling relations, 
as they are not subject to recent major group mergers, and should
represent archetypal old undisturbed systems. 

Recent studies of fossil groups show interesting features which
distinguish them from normal galaxy groups. 
\chandra observations of the nearest fossil galaxy group, NGC 6482, 
showed a very  high gravitational mass concentration and the absence of 
a cool core \citep{kjp04}, despite the short central cooling time.  
The \chandra and \xmm study of the fossil cluster,, RX J1416.4+2315
showed similar features \citep{kmpj06}. 
There has been only one statistical study of fossil groups \citep{jones03}, 
based on ROSAT pointed observations, which demonstrated some of the 
differences in fossils compared to normal galaxy groups and clusters.

The present study uses high resolution, short to moderate exposure,
\chandra observations of fossil groups drawn from this flux-limited
sample combined with previously studied fossils.
Together, this provides the largest sample of fossil galaxy groups studied
using the high resolution of \chandra.  Section 2 briefly
reviews fossil groups and describes the observations. The results from
imaging and spectral X-ray analysis and comments of individual systems
are presented in Section 3. The scaling relations involving X-ray and 
optical properties of fossils are presented in Section 4. Section 5 
studies the mass distribution, $M-T$ relation and mass-to-light ratio.
A discussion and concluding remarks can be found in section 6.

We adopt a cosmology with $H_0=70$ km~s$^{-1}$~Mpc$^{-1}$ and
$\Omega_m=0.3$ with cosmological constant $\Omega_\Lambda=0.7$
throughout this paper. 

\begin{figure}
\center
\epsfig{file=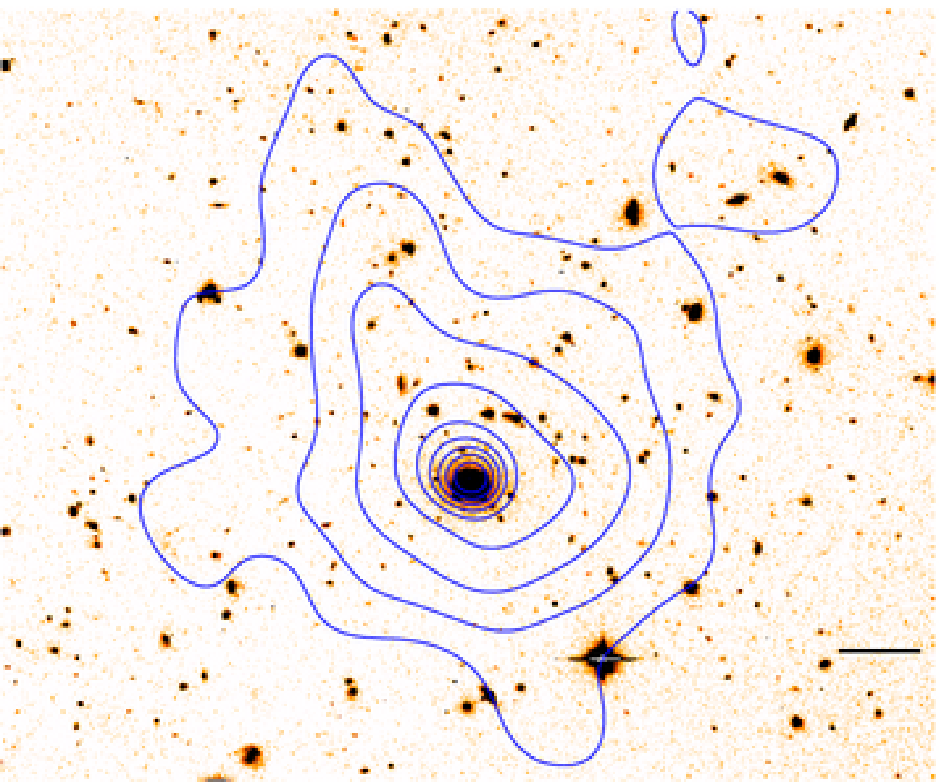,width=1.6in,height=1.4in}
\epsfig{file=beta1256.ps,width=1.6in,height=1.4in}
\epsfig{file=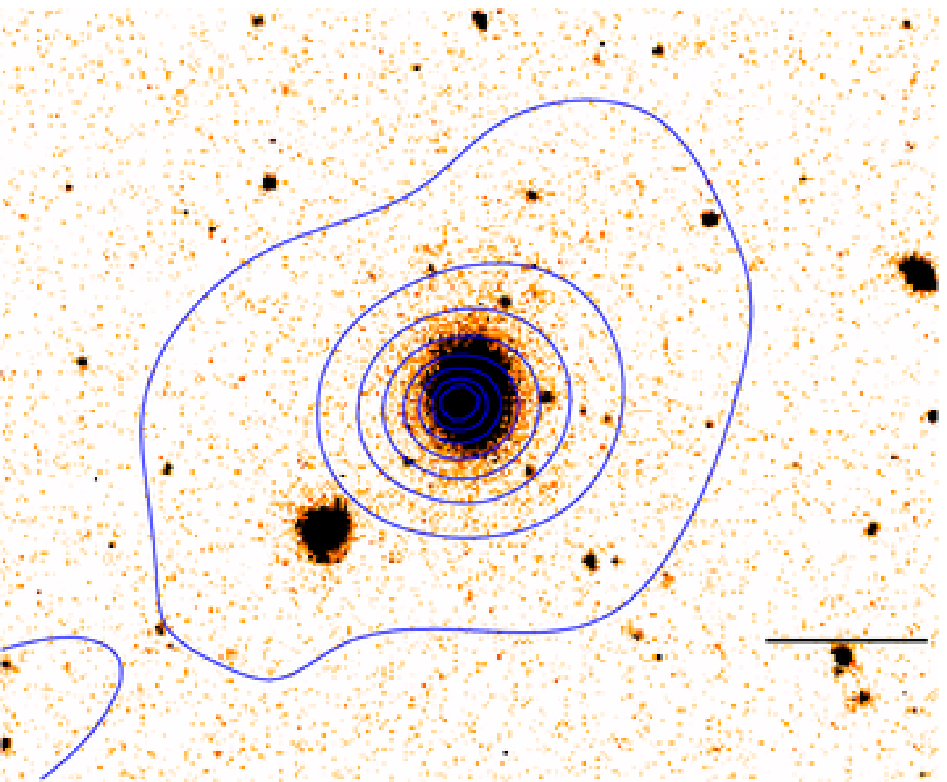,width=1.6in,height=1.4in}
\epsfig{file=beta1331.ps,width=1.6in,height=1.4in}
\epsfig{file=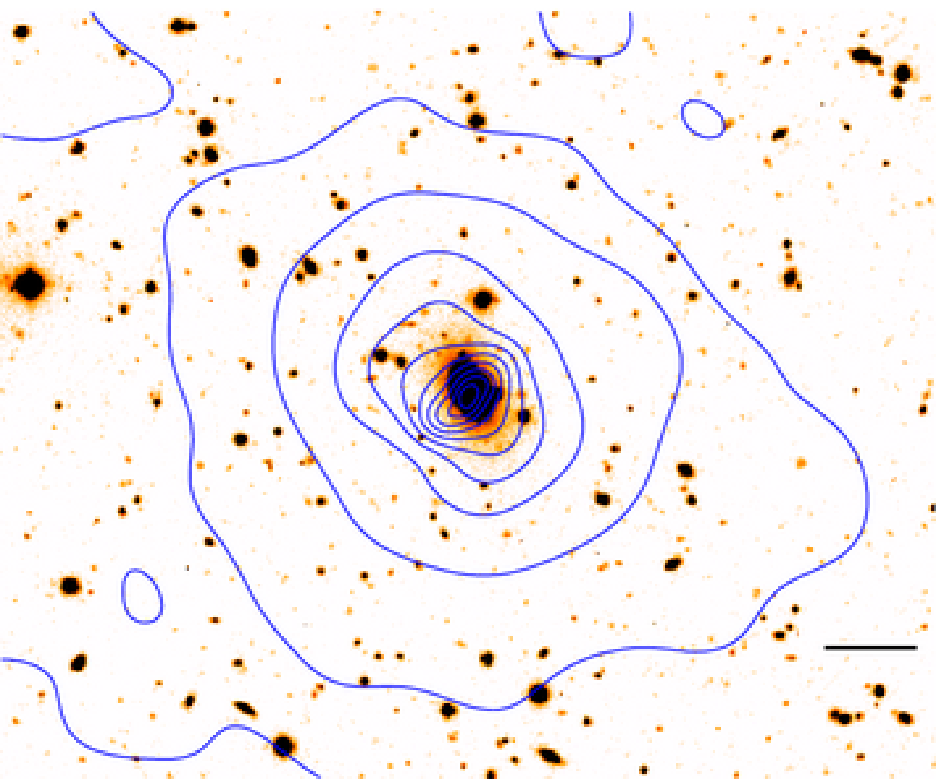,width=1.6in,height=1.4in}
\epsfig{file=beta1340.ps,width=1.6in,height=1.4in}
\epsfig{file=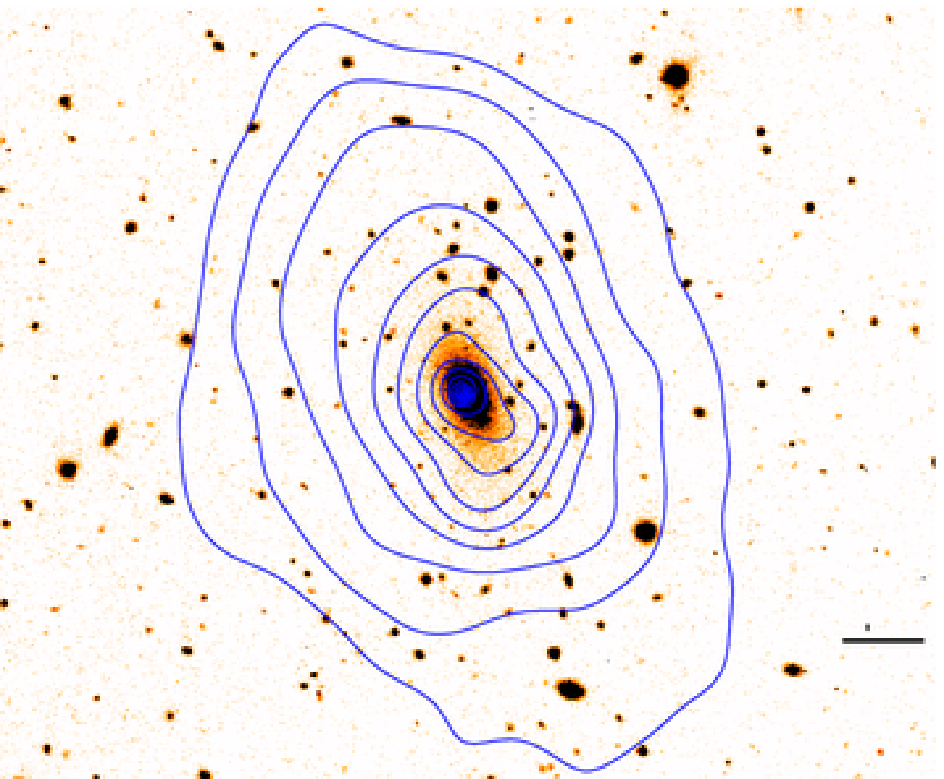,width=1.6in,height=1.4in}
\epsfig{file=beta1416-1.ps,width=1.6in,height=1.4in}
\epsfig{file=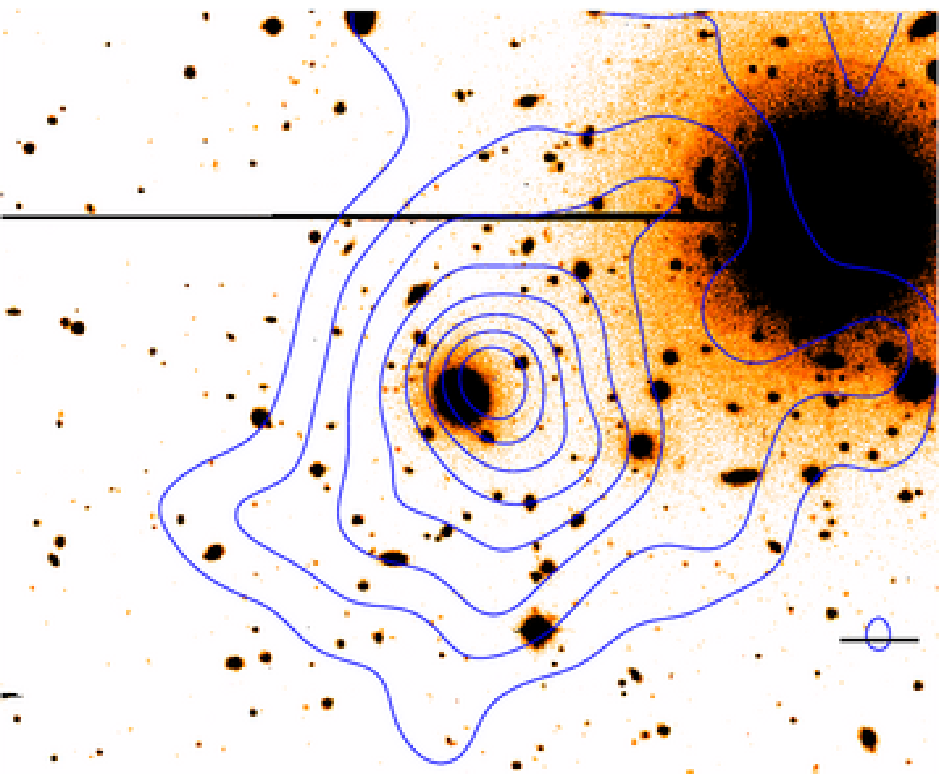,width=1.6in,height=1.4in}
\epsfig{file=beta1552.ps,width=1.6in,height=1.4in}
\caption{X-ray contours from the soft (0.3-1.0 keV) diffuse emission
overlaid on R-band INT optical images (left) along with radial X-ray surface 
brightness distribution and the best $\beta$-model fit (right) for 
RX J1256.0+2556, RX J1331.5+1108, RX J1340.5+4017, RX J1416.4+2315, 
RX J1552.2+2013 (top to bottom). The surface brightness profiles of the 
RX J1416.4+2315 is fitted with a double $\beta$-model \citep{kmpj06}.}
\label{chansoft}
\end{figure}

\begin{table*}
\begin{center}
\caption{Basic properties of the \chandra observations of fossil groups.  
\label{table1}
}
\begin{tabular}{llllcccl}
\hline
Group& R.A.  &   Dec. & z &t$_{obs}$ &  t$_{useful}$ & kpc/arcsec &Reference/Obs ID\\
     &(J2000)&(J2000) &   & ksec     &        ksec   &            &\\
\hline
RX J1256.0+2556 & 12:56:03.4 & +25:56:48 & 0.232 & 27.7 & 24.7& 3.73&3212\\ 
RX J1331.5+1108 & 13:31:30.2 & +11:08:04 & 0.081 & 30.7 & 27.6& 1.53&3213\\ 
RX J1340.5+4017 & 13:40:33.4 & +40:17:48 & 0.171 & 47.0 & 45.9& 2.92&3223\\
RX J1416.4+2315 & 14:16:26.9 & +23:15:32 & 0.137 & 14.6 & 14.4& 2.44&Khosroshahi \etal (2006)\\
RX J1552.2+2013 & 15:52:12.5 & +20:13:32 & 0.135 & 14.9 & 14.9& 2.40&3214\\\hline
NGC 6482    & 17:51:48.8 & +23:04:19 & 0.013 & 19.6 & 18.0& 0.27&Khosroshahi \etal (2004) \\
ESO 3060170  & 05:40:06.7 & -40:50:11 & 0.036 & 28.5 & 27.6& 0.71&Sun \etal (2004)\\

\hline
\end{tabular}
\end{center}
\end{table*}

\section{Fossil groups, the sample and observations}

Fossil galaxy groups are identified as galaxy groups dominated,
optically, by a single giant elliptical galaxy surrounded by an
extended X-ray emission with X-ray luminosity of $L_{X,bol} \geq
10^{42} h^{-2}_{50}$ erg s$^{-1}$. There is at least a 2 magnitude
difference between the luminosity of the first and the second ranked
galaxies, measured in the R-band, within the half virial radius. 
\citet{jones03} give the rationale for the above
choices. The lower limit to $L_X$ guarantees the
existence of a collapsed galaxy system, while the optical
criterion ensures that the M$\star$ galaxies have merged into a single
luminous galaxy as a result of dynamical friction \citep{binney87}.  
No upper limit is placed on the X-ray luminosity or
temperature, but as dynamical friction is the main player in
the formation of fossil groups, the mass and physical size of the
system cannot be arbitrarily large as the time scale for dynamical
friction to lead to orbital decay, and hence to galaxy merging, 
should not exceed the age of the universe \citep{kmpj06,milos06}. 

The central galaxy in fossils can be as luminous as the brightest cluster
galaxies (BCGs). However, despite apparent similarities, the isophotes of 
the central galaxy in fossil groups are found to be non-boxy in contrast
to the isophotes of BCGs which are predominantly boxy \citep{kpj06}.

\subsection{The X-ray sample}

Since the discovery of the first fossil \citep{ponman94}, similar
structures have been identified, sometimes branded as ``over luminous
elliptical galaxies'', or OLEGs \citep{vikh99,yoshioka04}, which meet
the fossil selection criteria \citep{jones03}. Over a dozen such system
have been identified in different surveys, or pointed observations,
and observed with a variety of X-ray instruments. 

The volume-limited sample of spatially extended X-ray sources compiled by
the WARPS project (Wide Angle ROSAT Pointed Survey; \citet{scharf97,
jones98, perlman02}) forms the core of the sample on which the present
study is based. Details of the search, initial sample
selection and subsequent observations leading to this flux limited
sample can be found in \citet{jones03}.  In order to improve our statistics,
we have added the results recent high quality
observations of fossils from the literature.  This includes the nearest
known fossil group, NGC 6482 \citep{kjp04}, and the massive fossil ESO
3607010 \citep{sun04}. In addition, results from the study of isolated
OLEGs by \citet{yoshioka04}, for which a mass analysis is available,
are also used in comparisons. We treat the latter sample as ``fossil
candidates'' excluding RX J0419.6+0225 (NGC 1550) which, according to
\citet{sun03}, does not satisfy the fossil optical selection criterion.

\subsection{\chandra observations} 

\chandra observations of the fossil groups in our sample
were performed in 2003 using \chandra ACIS-S. One major justification for
the \chandra observations of these systems was to obtain an accurate
position for any point sources, and hence to eliminate any
ambiguities regarding the point source contribution to the X-ray
luminosity of the fossils in the earlier study of \citet{jones03}. Also,
the earlier ROSAT observations were not of sufficient quality to give
a reliable estimate of the gas temperature, which is important for the study
of various scaling relations. The basic properties of our observed sample 
are summarised in Table 1 (top).  This includes 4 systems from the 
WARPS statistical sample \citep{jones03} plus the original fossil
\citep{jones00}, RX J1340.6+4018. Due to the limited statistics, we do not 
insist on limiting this study to the original flux-limited sample. Table 1 
also includes the basic information about two other fossils with existing 
\chandra data used in this study.

\subsection{Optical photometry and spectroscopy} 

The imaging and spectroscopic observations of the sample were performed
using the observational facilities of Issac-Newton Group of Telescopes
(ING), Kitt-peak National Observatory (KPNO). The R-band image was
obtained using the INT 2.5m wide field imager, with 20 min exposures
taken in service time. Unfortunately the conditions were not
photometric, and so further R-band imaging was obtained, in
photometric conditions, to calibrate the original images. For two
systems (RX J1416.4+2315 and RX J1552.2+2013) this further imaging was
obtained with the 8k mosaic camera at the University of Hawaii 2.2-m
telescope. For the other systems, the calibration images were obtained
in additional INT wide-field camera service time. The 
photometric accuracy for
all systems is $\le$0.05 mag. Absolute magnitudes of the central
galaxies were corrected for Galactic absorption. No K-corrections were
applied in deriving $\Delta m_{12}$, since the corrections are small
($\approx$0.1 mag) at the redshifts sampled, and in any case all of the
brightest galaxies, and most of the second-brightest galaxies, are of
an early type and would have identical K-corrections.

The spectroscopic observations were performed during a run to study
several fossil groups using a multi-slit spectrograph on the KPNO-4m
telescope in March 2000. The Ritchey-Cretien spectrograph
and KPC-10A grating gave a dispersion of 2.75 \AA/pixel over the
range 3800-8500 \AA, and with 1.8 arcsec slitlets a resolution of 6$\AA$
(FWHM) was achieved.  Risley prisms compensated for atmospheric
dispersion. Spectra were obtained through up to three slitmasks, with
typically an hour exposure on each.  The spectroscopic data were
reduced and analysed in the standard way using IRAF.

\section{X-ray analysis and results}

\subsection{Spatial distribution}

The \chandra data were first checked for bad pixels and flares using
the latest version of the standard analysis tools (CIAO 3.2 and CALDB
3.1).  To examine the diffuse X-ray emission,
point sources were detected using the CIAO ``wavedetect'' software, and
removed. Their locations were then filled with surrounding background values
using a linear interpolation. Figure \ref{chansoft} shows the contours of
soft (0.3-2.0 keV) diffuse X-ray emission overlaid on R-band images. In
general, the X-ray emission is relaxed, indicating an absence of any
recent merger. X-ray radial surface brightness profiles were
extracted from the the ACIS-S3 images in a soft energy range (0.3-2.0 keV) 
and fitted with a single $\beta$-model profile (see the Appendix) except for 
RX J1416.4+2315 where a double $\beta$-model was used \citep{kmpj06}. 
The data quality varies due to the distance and luminosity of the
targets and the exposure time. The binning is linear and each bin
has a minimum of 50 net counts, except RX J1552.2+2013, where this is
reduced to 20 net counts per bin. The results of the 
spatial analysis is given in Table 2. 
In all cases the centroid of the X-ray emission matches the centre of the
giant elliptical galaxy, except for RX J1552.2+2013 where an offset
of $\sim$5 arcsec is seen (see section 3.3.5).

\subsection{Spectral analysis}

We adopt the same analysis procedure as for NGC6482 \citep{kjp04}) and
RX J1416.4+2315 \citep{kmpj06}) for the targets in Table 1, except 
that we use the latest version of the analysis tools (CIAO 3.2 and CALDB 3.1).

With the high spatial resolution of \chandra, we were able to
exclude the point sources and examine the intra-group hot gas spectrum
without the presence of contaminating sources. Due to limited counts
we extract a global ACIS spectrum in all cases from a circular region 
within which the X-ray was detected. We fit an absorbed APEC
\citep{smith01} model with a hydrogen column density fixed to the
galactic value at the coordinates of each system. The background was 
chosen for each system from the same region on the re-projected blank 
sky observations to account for local variations within the chip. 

The results of the spectral analysis are given in Table 2. 
This table also lists the properties of other known fossils used in this 
study to achieve better statistics. Additional information on individual systems can be found in the appendix.

{\tiny
\begin{table*}
\begin{center}
\label{table2}
\begin{tabular}{lccccccccccc}
\hline
Group & $r_c$ & $\beta$ & $<T>$ & Abundances & $\frac{r_{spec}}{r_{500}}$& L$_X^a$
      & M$_{500}$ & $S_{core}^b$ &$tcool_{core}^c$  \\ 
Name & $kpc$ & & $keV$ & $Z_{\odot}$ & & $10^{42}ergs/s$ & 
$10^{12}M_\odot$&  $KeV~cm^2$& $10^9$yr \\
\hline
RX J1256.0+2556 &$64.1\pm31.7$ &$0.6\pm0.05$& $2.63\pm1.13$ & $0.58\pm0.55$ 
&0.64& 50 &86 &214&11 \\ 
RX J1331.5+1108 &$14.0\pm4.5$&$0.56\pm0.05$& $0.81\pm0.04$ & $0.36\pm0.27$
&0.44& 2.1&9.5&79&9.2 \\ 
RX J1340.5+4017    &$14.7\pm2.11$&$0.43\pm0.02$& $1.16\pm0.08$ & $0.16\pm0.05$
&0.61& 5.2&19&81&11\\
RX J1416.4+2315$^{d}$ & $129.6\pm9.6$&$0.54\pm0.02$ & $4.0\pm0.62$ & $0.23\pm0.07$ 
&0.97& 170&179&280&9.6\\
RX J1552.2+2013 &$117\pm70$ &$0.56\pm0.11$& $2.85\pm0.9$& 0.3 fixed
&0.22& 60&110&211&29\\\hline
NGC 6482$^{d}$  &$7.2\pm0.05$&$0.53\pm0.05$& $0.66\pm0.11$ & $0.6\pm0.2 $   
&0.16& 1.1&3.5&55&10 \\
ESO 3060170$^{d}$ &$47.9\pm 0.9$ &$0.53 \pm0.003$& $2.6\pm0.3$ & $0.5\pm 0.2$ 
&0.89& 66 &$98$&200&20\\\hline
NGC 1132$^e$ &- &-& $1.06\pm0.08$ & $0.19\pm 0.1$ &0.88& 2.1 &$6.8$&-&-\\
RXJ0454.8-1806$^e$&-&-& $2.14\pm0.17$ & $0.39\pm 0.3$ &0.73&13 &$49$&-&-\\
\hline
\end{tabular}
\end{center}
\caption {The X-ray properties of the fossil groups.
$^a$ The X-ray luminosity is bolometric and is measured within 
$r_{200}$.
$^{b,c}$ The entropy and cooling time are measured measured at $0.1r_{200}$.
$^{d}$ The X-ray surface brightness profile is better described with a double $\beta$-model fit.See References in section 2.1 for detailed X-ray surface brightness analysis. These systems have spatially-resolved temperature profiles. 
$^e$ A fossil candidate from the ASCA observations of OLEGs \citep{yoshioka04}}.
\end{table*}

}
\normalsize

\section{X-ray scaling relations}

\subsection{$L_X-L_R$ relation} 

One of the interesting properties of fossils  is the excess of about 
one order of magnitude in the X-ray luminosity of fossils compared to 
non-fossil groups for a given total optical luminosity \citep{jones03}. 
In the absence of spectroscopic data, the optical luminosity of the 
central galaxy was used for the $L_X-L_R$ comparison by Jones \etal (2003). 

Since then we have obtained the redshifts for a number of galaxies in
each group, and are now able to explore the $L_X-L_R$ relation with better
accuracy. To obtain the total optical luminosity (R-band) of fossils
we accumulate the luminosity of the spectroscopically confirmed member 
galaxies within $r_{200}$. The limiting magnitude of our 
spectroscopic observations is $M_R\approx-19$. Deeper galaxy 
luminosity functions (to $M_R\approx-17$) are available for 
RX J1416.4+2315 \citep{cyp06} and RX J1552.2+2013 \citep{mendes06}. 
We find that only 2-3\% of the total R-band luminosity of these 
groups lies in galaxies within the above two magnitude interval. 
This small correction is applied to RX J1416.4+2315, RX J1552.2+2013, 
RX J1256.0+2556 and RX J1331.5+1108. The contribution of 
the non-brightest group members to the total luminosity (in R-band) of our
fossil groups is estimated to be $\sim50-100$\% of the luminosity of 
the brightest galaxy. This same ratio is $\sim55$\% the in B-band, within 
$\approx0.5r_{200}$ for RX J1416.4+2315 and RX J1552.2+2013, 
\citep{cyp06,mendes06}. 

For the galaxy groups with very few members known -- RX
J1340.5+4017 and NGC 6482 -- we estimate the fraction of the missing light 
by comparing them to one of the above four systems,
for which more data are available, according to the luminosity of 
their dominant galaxy. We adopted this approach, instead of
estimating the total R-band luminosity of the faint end using
statistical background galaxy subtraction, because R-band data for RX
J1340.5+4017 covers only out to radius $\sim r_{200}$,
which limits the effectiveness of the statistical background
subtraction. In addition, the statistical subtraction is less
efficient in the absence of multi-colour data. The bolometric X-ray
luminosities of the groups within $\sim0.3r_{200}$ were measured
directly from their extracted spectrum. Given this, together
with the $\beta$-model parameters for each system (Table 2), we can estimate
the bolometric luminosities within any radius.
We choose $r_{200}$ for the comparisons under this subsection. The
error in the X-ray luminosity is obtained from the Poisson error in the 
X-ray count rate.

\citet{jones03} used a sample of X-ray bright groups from \citet{helsdon03} 
to make a comparison between fossils and non-fossils in the
$L_X-L_R$ plane. A larger and more representative sample of galaxy groups is
now available from \citet{osmond04}. We use the latter sample of groups, 
known as GEMS sample, for our comparisons through out this study.

\begin{figure}
\center
\epsfig{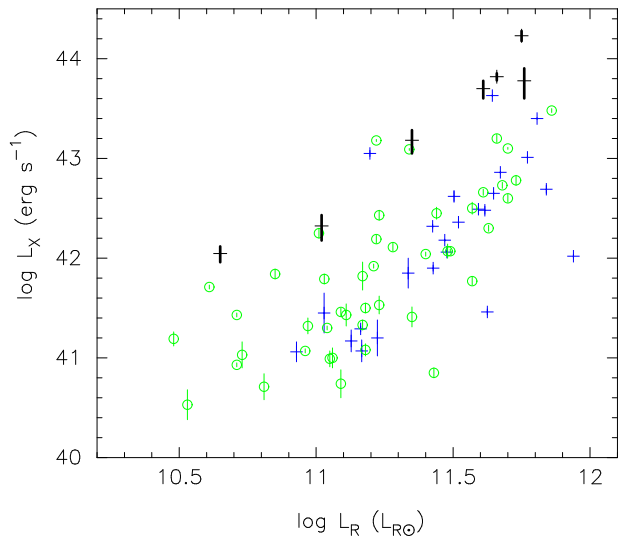}
\caption{$L_X$-$L_R$ relation: Comparison between fossils and non-fossil 
groups. The R-band luminosities are the total luminosity of the group.
Bold dark data points represent fossil groups in Table 2. GEMS 
groups \citep{osmond04} are shown with open (green) circles while 
thin crosses (blue) represent X-ray bright groups in \citet{helsdon03}. 
Fossils are more X-ray luminous for a given optical luminosity of the groups.}
\label{lxlopt}
\end{figure}

Figure \ref{lxlopt} confirms that fossils are more X-ray luminous than
non-fossil groups, for a given total R-band group luminosity.
The fossil systems lie at the extreme upper envelope of the distribution
for groups in general.

\subsection{$L_R-\sigma$ relation}

\begin{figure}
\center
\epsfig{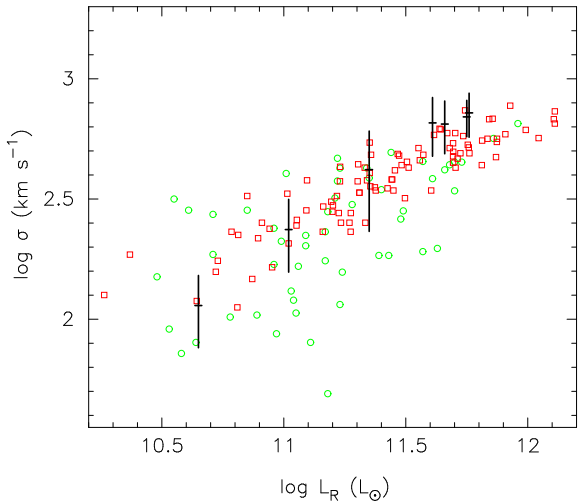}
\caption{$L_R-\sigma$ relation: Fossils are compared with non-fossil 
galaxy groups (GEMS) which are shown with open circles (green) and a
sample of groups and clusters from \citet{girardi02} shown with open
squares (red). It appears from this figure that fossils fall on the
distribution of groups and clusters.}
\label{loptsigma}
\end{figure}

One possible explanation for the difference in the distribution of
fossil and non-fossil groups in the above scaling relation (Fig
\ref{lxlopt}) is that fossils might be under-luminous in the optical,
due to inefficient star formation. If this were a strong effect then
there be nothing unusual about their X-ray luminosity. However, Fig
\ref{loptsigma} shows that any such effect is weak in clusters, and
non-existent in the group regime, where fossils fall on the
$L_R-\sigma$ relation of non-fossil galaxy groups. We will discuss these
features in Section 6 in the light of other scaling relations.

Group velocity dispersions are based on our spectroscopic observations 
of the sample in Table 1 and NED for the rest of the systems. They are 
calculated using the following relation, also used in the comparison
sample of \citet{osmond04}: 

\be
\sigma=\sqrt{\frac{\Sigma(v-\bar v)^2}{N-\frac{3}{2}}}\pm\frac{\sigma}
{\sqrt{2(N-3/2)}},~{\rm km~s}^{-1}.
\ee 

This estimator corrects for a statistical bias, which results if 
one uses the normal unbiased estimator for $\sigma^2$ and then takes
the square root to obtain $\sigma$ (which is then not unbiased). This
correction is the origin of the term 3/2 (rather than 1) in the denominator 
of the above equation.

\subsection{$L_X-T$  relation}

\begin{figure}
\center
\epsfig{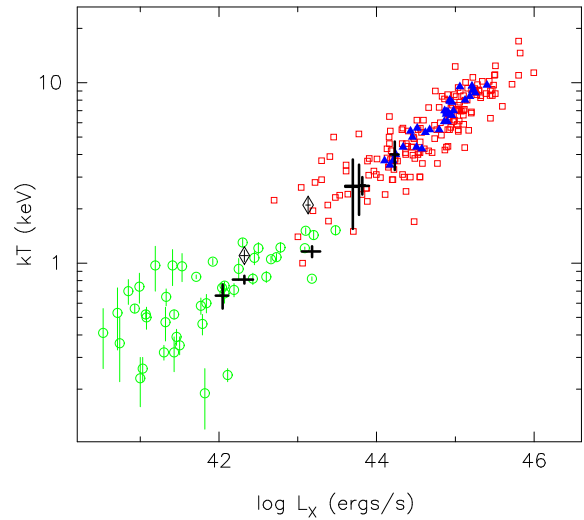}
\caption{$L_X$-$T_X$ relation: Comparison between fossils and non-fossil 
groups. The fossils data points consist of the data in Table 2 (dark bold) 
and the isolated OLEGs, as fossil candidates, in \citet{yoshioka04} are shown 
with diamonds. Fossils fall on the $L_X-T$ relation for non-fossil groups 
in the GEMS sample (open circles) and clusters. The cluster data are those of 
Markevitch (1998, filled triangles) and Wu, Xue \& Fang (1999, open squares). 
All the X-ray luminosities are bolometric.}
\label{LT}
\end{figure}

If excess X-ray luminosity in fossils were the only difference
between the X-ray properties of fossils and non-fossil groups, then they 
would be expected to deviate from the
$L_X-T$ relation known for non-fossil groups and clusters. This appeared to
be the case from an earlier ROSAT study \citep{jones03} based on very
limited statistics. Two fossil groups, for which \citet{jones03} could
measure the temperature, were found to have high X-ray luminosity
for their gas temperature. Based on this finding, it was argued that 
fossils are low-entropy systems due to their higher gas
density, in comparison to non-fossils. However we have shown in 
\citet{kmpj06} that the X-ray temperature of the RX J1416.4+2315 was 
underestimated in the \rosat analysis. The present wider study shows that 
the above 
system is not an exception and, as it is seen in Fig \ref{LT}, fossils 
fall on the conventional $L_X-T$ relation of non-fossil groups and clusters. 
Hence, if from our earlier arguments we assert that $L_X$ is enhanced in 
fossils, then it follows that they must also have elevated mean temperature
values, such that they remain on the standard group $L_X-T$ relation.

\subsection{$L_X-\sigma$ relation}

\citet{mahdavi01,xue00} have presented $L_X-\sigma$ relations for
clusters and groups. \citet{osmond04} found a slope of 2.31 $\pm$ 0.61
in $L_X-\sigma$ for their sample of galaxy groups with intergalactic X-ray
emission, flatter than the value of $4.5 \pm 1.1$ found by \citet{helsdon00}. 
There is a good deal of scatter in the relation, which may in part account for
the disagreement between various studies. While \citet{ponman96}, 
\citet{mz98}; \citet{helsdon00} and \citet{mahdavi01} find that groups 
are consistent with the
cluster-relation slope of $\approx 4$ , Mahdavi \etal (1997,2000) and
\citet{xue00} find significantly flatter relations in groups with a
slope similar to the finding of \citet{osmond04}.

\begin{figure}
\center
\epsfig{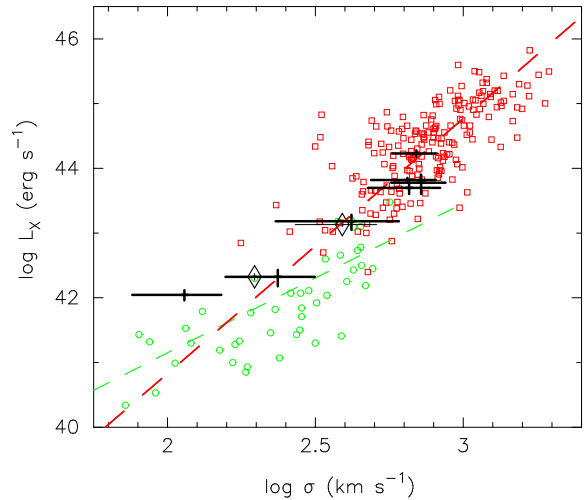}
\caption{$L_X-\sigma$ relation: Comparison between fossil and non-fossil 
groups. Fossils groups appear to be more X-ray luminous for a given group 
velocity dispersion. Symbols are the same as Fig \ref{LT}. The thick (red) 
and thin (green) dashed lines are the best fit to clusters and groups, 
respectively.}
\label{lxsigma}
\end{figure}

Figure \ref{lxsigma} shows the distribution of fossil groups
in the plane of $L_X-\sigma$ along with the non-fossil groups and
clusters. Fossils appear more X-ray luminous than non-fossil groups 
for a given group velocity dispersion. 
The slope of the relation for fossils is $2.74\pm0.45$, consistent with 
the slope of the relation for the groups in the GEMS sample. 
The flattening at $L_X\sim10^{42}$ erg s$^{-1}$ is due to the fossil 
X-ray selection criterion. We note that the statistics are 
limited and therefore it is not clear whether fossils are simply X-ray 
boosted version of galaxy groups (with a similar slope) or whether they 
follow the trend seen in galaxy clusters. 

\subsection{$T_X-\sigma$ relation}

Studies of the $T_X-\sigma$ relation in clusters show that the relationship
between velocity dispersion and gas temperature departs slightly from
the virial theorem expectation ($\sigma\propto T_X^{1/2}$) 
\citep{girardi98,wuxue99}. In the group regime, some studies have concluded
that groups follow almost a similar trend to
clusters (e.g. Mulchaey 2000; Xue \& Wu 2000), whilst others have found
that the relation steepens further at $T_X<1$ (Helsdon \& Ponman
2000). \citet{osmond04} show that there is a great
deal of non-statistical scatter in the groups, in addition to the
large statistical errors in both $T_X$ and especially $\sigma$, in the
poorest systems. This appears to be the origin of the controversy over
whether the relation does or does not steepen in the group regime.

\begin{figure}
\center
\epsfig{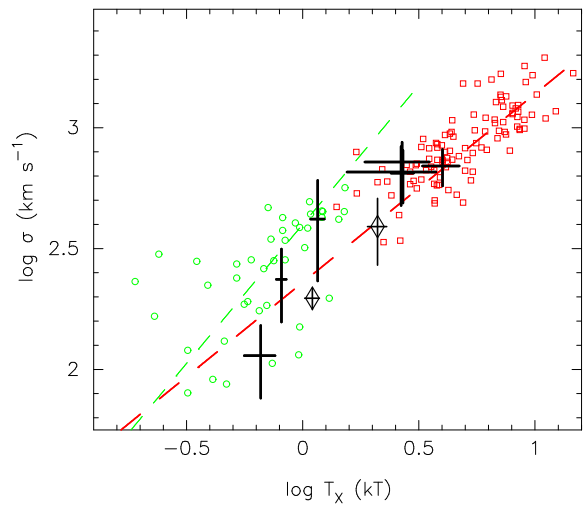}
\caption{$T_X-\sigma$ relation: Comparison between fossil and non-fossil 
groups. Fossils groups tend to be hotter for a given group velocity 
dispersion. Symbols are the same as Fig \ref{LT}. The thick (red) 
and thin (green) dashed lines are the best fit to clusters and groups, 
respectively.}

\label{txsigma}
\end{figure}

Figure \ref{txsigma} shows that fossil systems tend to be hotter than 
non-fossils in the group regime, as expected from the $L_X-T_X$ and 
$L_X-\sigma$ relations, and therefore consistent with a scenario in 
which the temperature 
and X-ray luminosity of fossils are both boosted relative to non-fossil 
systems.

The above two scaling relations linking the hot gas and galaxy dynamics
show significant departures of fossil groups from the scaling
laws of non-fossil groups. The fossil groups seem to scatter around an
extrapolation of the scaling laws derived for galaxy clusters, resulting
in an offset from the relations followed by normal galaxy groups.

\subsection{$S-T_X$ relation}

The gas entropy is defined here as $S=T/n_e^{2/3}$, where $n_e$ is the 
electron number density. At a given redshift, this quantity should scale 
linearly with temperature, owing to the self-similarity of the gas density 
profiles.

\begin{figure}
\center
\epsfig{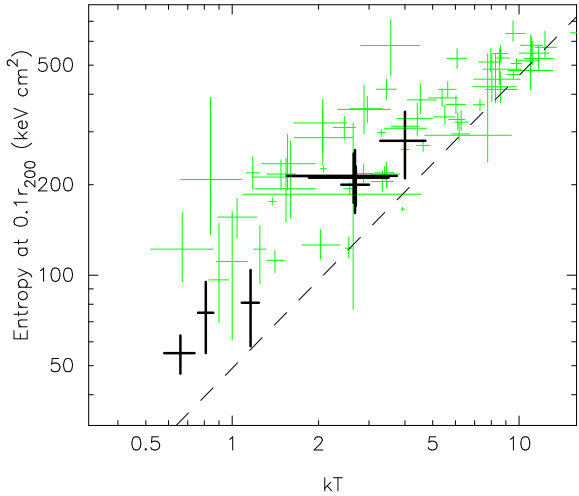}
\caption{Gas entropy at $0.1r_{200}$ as a function of system temperature for
fossils (dark symbols) and non-fossil clusters and groups in 
\citet{ponman03}. The dashed line is the expectation from self-similarity 
normalised to the mean entropy of the hottest clusters in \citet{ponman03}.}
\label{entropy}
\end{figure}

Estimating the gas entropy at $0.1r_{200}$ we compare in Fig
\ref{entropy} the $S-T_X$ scaling for fossils and non-fossils in
\citet{ponman03}. While there is still a significant overlap between 
fossils and non-fossils in this scaling relation, fossils lie along the lower envelope of the distribution. The effect is strongest in low mass fossils, 
which fall much closer to the scaling expected from self-similarity than do 
the bulk of normal groups. The entropy and its error was calculated 
using the method described in \citet{kjp04,kmpj06}.

The slope of the relation for the fossil systems in Fig~\ref{entropy}
is $\sim1.00\pm0.06$, consistent with slope of 1.0 expected from
self-similarity. As in \citet{ponman03}, we did not attempt to 
correct individual entropy values to the redshift of the system. Applying
such a correction is only appropriate if the system is observed at its
redshift of formation, which seems most unlikely in the case of fossils.
For reference, at $z\approx0.17$, such a correction (based on the
evolving critical density on the Universe) would
change the entropy by about $\sim$35\%.

\section{Mass distribution, concentration and the M-T relation}

The scaling relations involving the mass of groups and clusters
are amongst the most important ones. The mass distribution 
is also required for a precise measurement of overdensity radius.    

\subsection{X-ray derived mass}

The total gravitational mass at a given radius can be derived
from the gas density and the temperature profiles, assuming that the
gas is in hydrostatic equilibrium and is distributed with spherical
symmetry. 

The total gravitational mass is given by:
\be
M_{grav}(<r)=-\frac{kT(r)r}{G\mu m_p}[\frac
{dln\rho(r)}{dlnr}+\frac{dlnT(r)}{dlnr}].
\ee
where G and $m_p$ are the gravitational constant and proton mass, and
the mean particle mass is $\mu=0.6$.

We have already reported the detailed mass profiles of groups with 
sufficient counts to derive radial temperature profiles
\citep{kjp04,kmpj06}. Two other studies also give 
detailed mass analysis of fossils \citep{yoshioka04,sun04}. For the rest of 
the systems in Table 1 one could assume an isothermal temperature and derive
the mass using the above equation with a mean global 
temperature. Table 3 gives an estimate for the X-ray derived mass 
assuming such an isothermal model.

\subsection{Dynamical mass}

Having the group line-of-sight velocity dispersion, $\sigma$ (Table 3) we can
estimate the dynamical mass, which gives an independent measure of the
underlying total mass of the group. We calculate a virial mass using
the standard equation given by \citet{ramella89},
$M_{Dyn}=6\frac{r_{vir} \sigma^2}{G}$, where $r_{vir}$ is the
estimated or measured (see below) virial radius of the system. The 
virial dynamical masses, $M_{vir}$, are given in table 3.

\subsection{Overdensity radius}

Measuring the overdensity radius requires a 
detailed knowledge of mass distribution. In the absence of high quality 
data for some of the groups in this study we have to rely on an estimate. 
Some of the well known estimators of $r_{200}$ and $r_{500}$ are derived 
from simulations, for instance \citep{evrard96}: 

\be
r_{500}=1.77(\frac{kT}{10 keV})^{1/2}  Mpc. 
\ee

Using the definition of $\beta_{spec}=\frac{\mu\sigma^2}{kT}$, 
the above can be rewritten in terms of $\sigma$
\be
r_{500}=\frac{0.098\sigma}{H_0\sqrt{\beta_{spec}}} Mpc.
\ee 

Alternatively, one could use the virial theorem, directly, to estimate
the virial radius ($r_{vir}$), and any overdensity radius for a given
mass profile. \citet{girardi98} derives an approximation for the virial 
radius of a spherical system: 
 
\be
r_{vir}=\frac{0.2\sigma}{H_0} Mpc. 
\ee

$r_{500}$ can then be calculated using the definition of the density
contrast and by assuming a simple mass profile. This results in
$r_{200}\approx 1.58r_{500}$ and $r_{vir}\approx1.66r_{500}$, assuming
$r_{vir}\approx r_{180}$. This means that values of $r_{500}$
calculated using the latter indicator (eq. 6) are expected to be about
25\% higher in comparison to those from equation (4), for a system
with $\beta_{spec}=1$.

Recently \citet{willis05} compared values of $r_{500}$ derived 
from an isothermal model  to  
corresponding values obtained using the observed relationship between 
cluster mass and temperature. Employing the data presented by 
\citet{fin01} they obtain the following relation for 
a wide range of temperatures (0.75-14 keV),

\be
r_{500}=0.39 T_X^{0.63} h_{70}(z)^{-1} Mpc. 
\ee

Table 3 compares the values $r_{500}$ for our systems using the various 
estimates described above. Table 3 shows that measured values
of $r_{500}$, for RX J1416.4+2315, NGC 6482 and ESO 3060170, 
are systematically smaller than the values derived from an isothermal
hydrostatic model, but are in reasonable agreement with
values estimated using equation (6), within 5\% to 20\%. 

{\tiny
\begin{table*}
\begin{center}
\begin{tabular}{lcccccccc}
\hline

Group & $N_{spec}$& $\sigma$ &$\beta_{spec}$&$r_{500}$&$r_{500}$&$r_{500}$&$r_{500}$& $M_{vir}^{dyn}$\\ 
Name  &           &km s$^{-1}$ &    &   Mpc & Mpc ($iso$) & Mpc ($Eq.~ 6$) & Mpc ($Eq.~ 4$)&$10^{12} M_\odot$\\
\hline
RX J1256.0+2556  &8  & 773$\pm$214  &1.42&  -    & 0.72 &0.58 & 0.91 & 710  \\ 
RX J1331.5+1108  &6  & 236$\pm79$   &0.57&  -    & 0.36 &0.26 & 0.44 & 24   \\ 
RX J1340.5+4017  &4  & 419$\pm$187  &0.94&  -    & 0.44 &0.36 & 0.61 & 130  \\
RX J1416.4+2315  &18 & 694$\pm$120  &0.75& 0.77  & 0.88 &0.82 & 1.12 & 656  \\
RX J1552.2+2013  &13 & 721$\pm$150  &1.13&  -    & 0.76 &0.67 & 0.95 & 640  \\
NGC 6482         &5  & 115$\pm$38   &0.13& 0.25  & 0.34 &0.30 & 0.45 & 6   \\ 
ESO 3060170      &15 & 648$\pm$160  &1.01& 0.62  & 0.74 &0.67 & 0.90 & 469  \\ 

\hline
\end{tabular}
\end{center}
\caption{Dynamical properties of fossil groups}
\end{table*}
}
\normalsize

\subsection{M-T relation}

\begin{figure}
\center
\epsfig{file=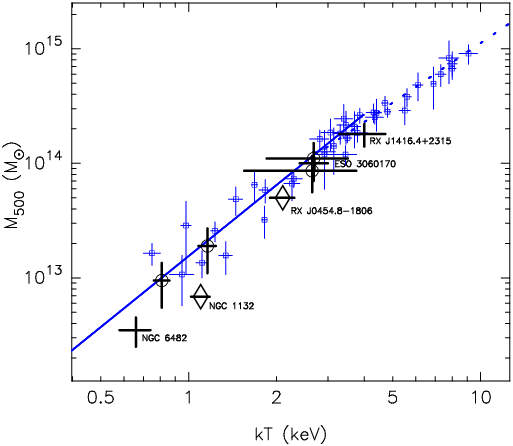,width=3in}
\caption{$M-T$ relation for fossils: Bold crosses show fossils listed in Table
2. Additional circle indicates the isothermality assumption in deriving the 
mass. Such an assumption results in over-estimation of the total mass 
\citep{sand03} . Two diamonds represent OLEGs 
\citep{yoshioka04}. Fossil groups 
seem to have higher temperatures, for their masses, in comparison 
to the non-fossil systems \citep{fin01} shown with open squares. 
The thick solid line (blue) is the best
orthogonal fit to the galaxy systems with $kT \le 4$. The dashed 
line (blue) is the same for galaxy systems with $kT \ge 3$.}
\label{MT}
\end{figure}

Three of the systems, RX J1416.4+2315, NGC 6482 and ESO 3060170, had
their mass profiles (and their associated errors) measured previously,
therefore we use their measured total mass within $r_{500}$ in Fig
\ref{MT}. The NFW fit parameters given in \citet{yoshioka04} were used
to estimate the $M_{500}$ for OLEGs, which are shown with diamonds in
Fig \ref{MT}. For the rest of the groups in the sample (Table 2) we
use the $r_{500}$, and corresponding value of $M_{500}$, derived from
equation (6) above. For comparison to non-fossil groups we use the
best orthogonal fit, $M_{500}=1.55(\pm0.25) 10^{13} M_\odot
kT^{2.06\pm0.16}$ from Willis \etal (2005) to the galaxy systems with
$kT<4.0$ keV studied by \citet{fin01}. As the figure shows, fossils
with resolved temperature profiles show a tendency toward higher gas
temperature for a given system mass -- an effect which is more
noticeable in low mass systems. All the fossils with resolved temperature 
profiles lie below the best fit M-T relation, while isothermality 
assumption for the rest of the fossils allow them to be consistent with 
non-fossils best fit relation. Given that isothermality assumption results 
in relatively higher masses \citep{sand03}, therefore we argue that 
fossils are, in general, hotter than non-fossils for a given mass.  The
error in the mass of fossils with unresolved temperature profile was
obtained from the relative error in the mass, using equation 2, based
on the errors in the surface brightness profiles and the global
temperature.

A comparison to the $M_{500}-T_X$ relation for $>2 keV$ clusters
derived by \citet{arnaud05} also shows that fossils are hotter for given mass. In a more recent study \citet{vikhlinin06} find a higher normalisation 
for the $M_{500}-T_X$ relation which they argue to be the consequence 
of the density profile steepening at large radii, which leads to their masses
being larger than would be obtained from a $\beta$-model representation
of the density profile. Their results are not strictly comparable to our 
sample, as they deal with hotter systems. We discuss the effects of cool core
correction, and the size of the area from which the global spectrum
was extracted, on the scaling relations in the discussion.  We discuss
the effects of cool core correction, and the size of the area from
which the global spectrum was extracted, on the scaling relations in
the discussion.

The fossils in this study cover a wide range of redshift, however, the 
effect of evolution in the $M-T_X$ relation (i.e. the $E(z)$ term in 
\citet{maughan06}) is smaller than the errors associated with the 
estimate of the mass for each system. A typical decrease in the mass
for a system at $z\sim0.2$ is only $\sim$7\%, relative to the best-fit 
$M-T_X$ relation for the non-fossil systems, which lie at $0.01\le z \le0.08$.
The effect will be negligible in the lowest
temperature fossils, which lie within $z\sim0.08$.

\subsection{Mass concentration}

Dark matter halos with an early formation epoch tend to be more
concentrated \citep{nfw95}. Recent numerical studies also predict
some mass dependency for the halo concentration, resulting from
the fact that lower mass halos generally form earlier
\citep{bullock01,dolag04}. Observational results seem to agree well
with the numerical predictions \citep{pratt05,pin05,vikhlinin06}.

We have previously reported a very high mass concentration,
$c_{200}\sim60$,
in the low mass fossil group NGC 6482 \citep{kjp04}. A recent study by
\citet{phil06} finds high values of mass concentration for a
sample of early type galaxies, after accounting for the baryons, however
only NGC 6482 qualifies as a fossil, all the other low mass systems in
this sample being too X-ray dim.

More recently we found a mass concentration of $c_{200}\sim12.5$
in the fossil cluster, RX J1416.4+2315\citep{kmpj06}, which is also
high for a $\sim 10^{14}$ $M_\odot$ cluster. In the case of this
massive fossil, our study showed that the stellar mass is not a
major contributor, and the mass concentration reduces only slightly, to
$c_{200}\sim11.2\pm4.5$ when the stellar contribution accounted for.
Fitting an NFW profile to the total mass distribution, \citet{sun04} found
a concentration of $8.5$ for the fossil system
ESO 3060170. While its possible that the
NFW profile is not the best model to describe the dark matter distribution
in fossils, all the existing
data seem to indicate that the total mass concentration in fossils is
higher than that observed in non-fossil systems of similar mass
\citep{pratt05}. 

\begin{figure}
\center
\epsfig{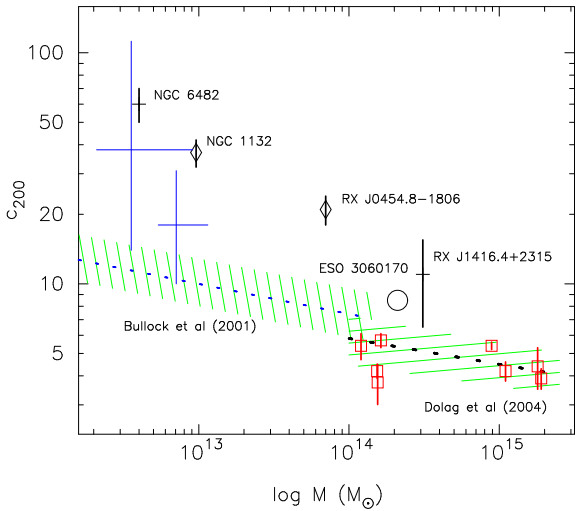}
\caption{$M-c_{200}$ relation: Comparison between the mass concentration 
in fossils and non-fossil groups and clusters. Three fossils with  
resolved temperature profile and two isolated OLEGs (diamonds) are 
compared with non-fossil clusters (open squares) from \citet{pratt05}. 
The expected values of the dark matter concentration 
and its variation with the halo mass from the numerical studies of 
\citet{dolag04} and \citet{bullock01} also are presented. The reasonable 
agreement between the non-fossils mass concentration from \citet{pratt05} 
with the dark matter concentration from the study of \citet{dolag04}
suggest that the excess concentration in fossils is due to their dark matter 
concentrations. In low mass end, two recent 
estimates (90\% confidence) for the concentration of 
NGC 6482, after accounting for the baryonic matter (see \citet{phil06} 
for details) are also shown (blue) which again lie above the numerical 
expectations.}
\label{c200}
\end{figure}

Fig \ref{c200} shows the variation of the mass concentration,
$c_{200}$, for fossils and non-fossils \citep{pratt05} with the total
halo mass.  It also shows the expected mass concentration of dark
matter halos from numerical models \citep{bullock01,dolag04}. In
general the mass concentration of fossil lie above the
expectations. In their recent study of relaxed clusters
\citet{vikhlinin06} find slightly higher values of mass concentration,
when fitting and NFW profile to the total density excluding the region
associated with the central brightest cluster galaxy
$r<$0.05$r_{500}$, but the difference is not as pronounced as in
fossils. They argue this to be due to the selection effect as they
focus on relaxed clusters and/or the radiative cooling of baryons and
the associated galaxy formation. They estimate the effect to be
$\Delta c_{500}\approx 1.0$ in clusters which is equivalent of $\Delta
c_{200}\approx 1.5$. It appears that the excess concentration seen in
fossil systems is generally more than it can be explained by the
\citet{vikhlinin06} arguments.

Distinction has to be made between the mass concentration and the dark
matter concentration. As we see in Fig \ref{c200} in the more massive
systems the difference between the two is reduced due to relatively
small contribution from the stellar mass. The fossils in this figure
are directly comparable to the non-fossil values of \citet{pratt05}, 
for a given mass, as they both refer to mass concentration. Given the
good overlap between the values of \citet{pratt05} and the predicted
dark matter concentration \citep{dolag04}, we argue that the 
offset is mainly due to higher dark matter concentration in fossils.

\subsection{Mass-to-light ratio}

There have been conflicting findings on the mass-to-light ratio of
fossil groups. While \citet{vikh99} report a high mass to light ratio
for OLEGs, we found a normal ratio for the low mass fossil NGC 6482
\citep{kjp04}. The value reported for ESO 3060170 is not high
either. The M/L values reported for two of the systems in
\citet{yoshioka04} are very high, but this is not surprising considering
that these authors added only 10-20\% to the luminosity of the
central galaxy to account for the luminosity of the rest of the
galaxies in the group. Recent studies of the galaxy luminosity
function in fossils
\citep{kmpj06,mendes06} show that non-brightest galaxies 
contribute at least as much as the brightest galaxy to the total luminosity
of the group. This can lower the value of the mass-to-light ratio by a
significant factor.

The method used here to estimate the total R-band luminosity of fossils 
is described in section 4.1. To estimate the B-band luminosity 
we can assume B-R=1.5 for the member galaxies, which would be appropriate
if all were early-type galaxies. In practice, this assumption results in an 
underestimation of the total B-band luminosity, and hence give an upper limit 
for the B-band mass-to-light ratio of fossils. 

Lower and higher values of M/L are also reported for galaxy clusters 
and groups. There are relatively large errors in the estimation of M/L 
because of uncertainties in the total masses, and the difficulties in 
obtaining a precise estimate of the optical 
luminosity of groups. For example, there is a factor of 2 difference 
between the aperture luminosities quoted in Girardi \etal (2002), based 
on APM and COSMOS data. Sanderson and Ponman (2003) evaluated a logarithmic 
mean value for M/L$_B$, of $243^{+33}_{-29}h_{70}$, substantially lower 
than the ordinary, arithmetic mean value of $318 \pm 52$. 

\begin{figure}
\center
\epsfig{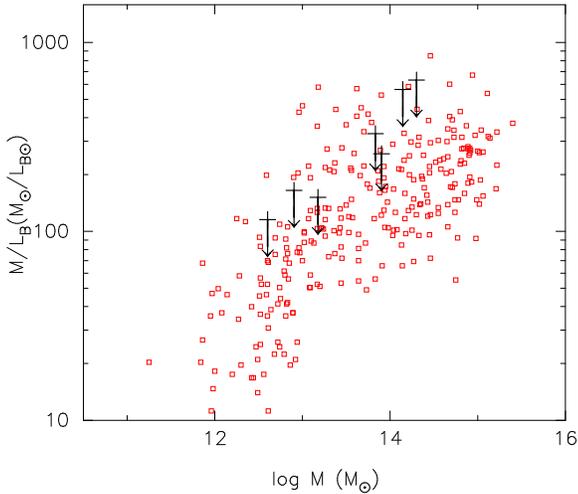}
\caption{Variation of B-band mass-to-light ratios of fossils and
non-fossil groups and clusters with cluster mass. The crosses (+) are the
data points from this study assuming that all the galaxies in fossils are 
early-types (B-R=1.5). Knowing that the dominant galaxy in fossils is a giant
elliptical, arrows give the amount of correction to be applied 
if we assume that all the non-brightest group galaxies are late-types.
The non-fossil groups and clusters data points (red squares) are from
\citet{girardi02}. Fossils tend to have high mass-to-light ratios, lying
along the upper envelope of the distribution, but they are not 
altogether exceptional.}
\label{moverl}
\end{figure}

Fig \ref{moverl} shows the M/L$_B$ for galaxy clusters, rich and poor,
from the study by Girardi \etal (2002) and the same for fossil galaxy
groups in our sample within the virial radius. We give two
estimates for the fossils. Firstly assuming that all the groups
members are early-type galaxies with (B-R=1.5), and secondly assuming
that all the groups members, except the dominant galaxy, are
late-types (B-R=0.8). This gives a lower limit for M/L in the
B-band. Based on our estimate of the contribution from the non-central
galaxy to the total luminosity of the group (section 4.1) and the
above argument, the mass-to-light of OLEGs studied by
\citet{yoshioka04} should also be reduced typically by half and would
therefore overlap with the range of the M/L presented for fossils in
this study.

\section{Discussion and Conclusions}

We have presented various X-ray related scaling relations for the largest
sample of fossil groups with high resolution X-ray data from 
\chandra observations. The main results 
are as follows:

\begin{itemize}

\item Fossils are more X-ray luminous for a given total optical 
luminosity of the group across the full mass range.

\item No noticeable difference is found in the $L_X-T_X$ relation of 
fossils compared to non-fossil groups and clusters.

\item There is no systematic difference in the plane of $L_{opt}-\sigma$ 
between fossils and non-fossil groups and clusters.

\item For a given group velocity dispersion, fossils are hotter and more 
X-ray luminous than non-fossil groups in the group regime. However, they 
tend to follow galaxy clusters
in the $L_X-\sigma$ and $T_X-\sigma$ planes. 

\item The scaling of the gas entropy with the temperature in fossils 
resembles a self-similar scaling -- the cooler fossils show much less excess
entropy than normal galaxy groups.

\item Total gravitational mass is more centrally concentrated in 
fossil groups and clusters compared to non-fossil systems.

\item The $M_X-T_X$ relation suggests that fossil groups are somewhat
hotter than non-fossil systems, for a given total mass of the group.

\item The mass-to-light ratio of fossils falls in the upper envelope
of the values seen in normal groups and clusters.

\end{itemize}

The first four of these results indicate that although fossil groups
fall on the same $L_X-T_X$ relation as normal groups and clusters, this results
from enhancement in both $L_X$ and $T_X$, compared to normal groups, for
systems of a given mass or optical luminosity. This difference in properties
could in principle arise in two different ways. It could be driven by
differences in the gravitational potential of fossils, or from a distinction
in the properties of the hot gas (within a similar potential). Or, of course,
both factors could contribute.

In favour of the first option, we have the observed evidence that the
potential is more cuspy (i.e. has higher concentration parameter) in
fossils (Fig.\ref{c200}). This will lead to greater compression of gas
in the inner regions of the system, and hence to both higher $T_X$
and enhanced X-ray emissivity. However, such a situation might be expected
to lead to enhanced cooling, which would in turn
raise the entropy of the gas in the inner regions, as a result either
of the removal of low entropy gas, or due to triggering of feedback
from AGN or star formation (Voit \& Bryan, 2001).
Given the observation (Fig.\ref{entropy}) that the entropy in fossil
groups is actually {\it lower} than that in other groups, it seems unlikely
that the difference in gravitational potential can be the whole story.

A second option, suggested by the result that the entropy in fossils
is close to a self-similar extrapolation from high mass clusters, with
much less sign of excess entropy than is seen in typical groups, is that
the mechanisms which raise the entropy of the intergalactic medium (IGM) 
have functioned much less effectively within fossil systems.
\citet{ponman03} pointed out that the large excess entropy seen outside
the core of groups was unlikely to be achieved by feedback operating
after virialisation, and \citet{voit03} and \citet{ponman03} suggested
that instead, most feedback operates within the IGM before it
falls into groups, reducing density contrasts in the pre-collapse
gas and leading to higher post-shock entropy once the gas crosses
the accretion shock during group formation. Within this scenario,
the lower entropy seen in fossil groups could be a natural consequence
of their very early formation, since feedback from supernovae or AGN
within galaxies would have less time to operate before the gas is
incorporated into the virialising group. Subsequent cooling could still
lead to a rise in entropy, but is now believed \citep{voit05,peterson03}  
to be limited by feedback from central AGN within groups and clusters.

Table 2 shows that for most of the systems the regions used for the 
  spectral analysis cover a reasonable fraction of $r_{500}$. 
  There are however two cases where the spectral extraction
  regions are small. NGC 6482 is probably the most critical one because
  of its low mass. The detailed study of the target \citep{kjp04} shows
  no sign of a cool core in this fossil. Therefore no correction for the
  cool core was needed. In fact the other fossils with resolved
  temperature profiles show no sign of a cool core either. ESO 3060170
  has a cool core of about 10 kpc
  \citep{sun04} much less than the typical $0.15r_{500}$ assumed for the 
  cool core exclusion. Similarly RX J1416.4+2315 gives no indication of 
  the cool
  core presence \citep{kmpj06}. This could be one reason why the mean
  temperature of fossils appears boosted. It is not clear if this
  is the work of AGNs, as they are present in almost all central cluster
  galaxies regardless of their status as fossils or non-fossils.

In summary, then, we suggest that the difference in the gas properties
of fossil systems, especially in the group regime, results in two
ways from their early formation epoch. Firstly the cuspy potential tends to
raise the luminosity and temperature of the IGM, however this works
in conjunction with the lower gas entropy (especially in lower mass
systems) compared to normal groups, which may arise from less effective
preheating of the gas in these early-forming systems. The rather high
mass-to-light ratios seen in fossils (Fig.\ref{moverl}) also suggests
that galaxy formation efficiency may be rather low in these systems,
another likely result of the early shock heating of their IGM.

\section{Acknowledgements}
We would like to thank the Issac Newton Group of telescopes for the
INT service observations, John Mulchaey for his involvement in the
spectroscopic observations and the anonymous referee for constructive
comments which helped to improve the content and the presentation of
the paper. We would like to thank Arif Babul, Ben Maughan and Ewan 
O'Sullivan for useful discussions and
comments. Alastair Sanderson has kindly provided the entropy and total mass 
estimates for comparison. This research has made 
use of the NASA/IPAC Extragalactic
Database (NED) which is operated by the Jet Propulsion Laboratory,
California Institute of Technology, under contract with the National
Aeronautics and Space Administration.

\section{Appendix - Notes on individual systems}

\subsection{RX J1256.0+2556} 

There has been an uncertainty in the formal classification of this system 
as a fossil \citep{jones03} because of the uncertainties in the
temperature, and hence in the virial radius.  The \chandra
observation still does not provide enough photon counts for a detailed
analysis of this system, but the spectral fit gives a mean temperature
within the central 100 arcsec. With the current estimate of the
temperature, $2.63\pm 1.13$ keV, the virial radius is $\sim 1.1\pm0.3$ 
the optical criterion $\Delta m_{12}\ge 2.0$ mag is not met, as
there exists one galaxy, spectroscopically confirmed, at a projected
distance of 550 kpc from the central galaxies with $\Delta m_{12}\sim
1.5$. A recent XMM-Newton analysis of this system \citep{mulch06} 
gives a hot gas temperature of $2.5^{+0.8}_{-0.6}$ keV. Given the large 
uncertainty in the temperature, the group could be a fossil. Thus 
we include it in our analysis and treat it as a 
fossil. We find no noticeable difference between this system and 
other confirmed fossils in its X-ray and optical properties.

\subsection{RX J1331.5+1108} 

There was an uncertainty in the X-ray luminosity of diffuse, hot gas
in this source in the earlier study \citep{jones03} because of the
possibility of point-source contamination. We measure the total X-ray
luminosity within a 75 arcsec radius region, from which the spectrum was
extracted, to be $L_X=0.71\times10^{42}$ (0.3-8.0 keV). The bolometric
luminosity from within $r_{200}$ (see section 5.3 for the estimation
$r_{200}$), from extrapolation, is $L_{r200}=1.2\times10^{42}$. We
therefore confirm that this system meets the fossil criteria based on
its X-ray properties. The spectroscopic membership observations
show that the optical criteria is also met and therefore the system is
a fossil group.

The central galaxy, at the X-ray peak, has narrow H$\alpha$ and [SII]
emission lines. An unresolved ($r< 2$ arcsec) $12.6 \pm 0.6$ mJy radio
source at 1.4 GHz is coincident with the central galaxy (from the
FIRST survey; White \etal 1997). This flux corresponds to a power of
$4.1 \times 10^{30}$ erg s$^{-1}$ Hz$^{-1}$ at 1.4 GHz, similar to
that of the first fossil system found by \citet{ponman94}  and
\citet{jones00}, and comparable to values found for radio-loud cD
galaxies in cluster cores. 

\subsection{RX J1340.5+4017} This is the original fossil galaxy group
studied by \citet{jones00} in detail. The optical properties,
including deep R-band photometry and spectroscopy of the central galaxy, 
are discussed in \citet{jones00}. We use a circular region of 
75 arcsec radius for the global spectrum 
extraction. A more detailed analysis of this system,
combining optical and X-ray observations is underway.
The relatively longer observation of the target shows significant 
asymmetry in the X-ray emission of the group in the core. The X-ray 
emission is elongated and is aligned with the central galaxy. 

\subsection{RX J1416.4+2315} This is the most massive and hot 
fossil system known, and is effectively a fossil{\it cluster}. Details of 
its X-ray properties from combined \chandra and \xmm observations, as well 
as optical photometry and spectroscopy of this galaxy cluster can be 
found in \citet{kmpj06}. A galaxy luminosity function of the system 
has been derived by \citet{cyp06}.

\subsection{RX J1552.2+2013} The X-ray observation of this target is 
relatively short, but the X-ray emission is clearly extended.  The 
background, bright QSO point source $\sim$2.5 arcmin west of 
RX J1552.2+2013, at z= 0.25, was excluded. A 60 arcsec radius region
was used for the global spectral analysis.

Three point X-ray sources were detected in the core of the system
matching the position of the central giant elliptical galaxy and two
other galaxies belonging to the group.  This is a rare incident as the
three seem to be perfectly aligned and are separated by an equal
distance of about 22 arcsec from each other. There appears to be an
offset of about 5 arcsec (see Fig \ref{chansoft}) between the centroid of
the X-ray emission and the the centre of the giant elliptical galaxy
in this fossil. This offset is unlikely to be real. The optical astrometry
should be good to 1 arcsec, but the determination of the X-ray centroid is 
significantly worse, due to the limited counts, and additional 
uncertainties arising from the removal of an
X-ray point source associated with the central galaxy. A galaxy
luminosity function for this fossil system is given in \citet{mendes06}.

\bsp
\label{lastpage}

%\newpage
%\begin{figure}
%\center
%\epsfig{file=MTsand.ps,width=3in}
%\caption{$M-T$ relation for fossils compared with the $M-T$ relation of non-fossil 
%clusters studied by Sanderson et al 2003. Bold crosses show fossils listed in Table
%2. Two diamonds represent OLEGs \citep{yoshioka04}. Fossil groups 
%seem to have higher temperatures, for their masses, in comparison 
%to the non-fossil systems studied by Sanderson et al 2003. 
%shown with open squares. The solid line is the best fit to the non-fossils.}
%\label{MTsand}
%\end{figure}

\end{document}